\documentclass{JHEP} 
\usepackage{epsfig} 
 
\title{Inflation After Preheating} 
 
\author{Gary Felder\\  
    Department of Physics, Stanford University, Stanford, CA
94305, USA\\ and  CITA, University of Toronto, 60 St George Str, Toronto, 
ON M5S 1A1, Canada\\ 
    E-mail: \email{gfelder@leland.stanford.edu}}
\author{Lev Kofman\\  
    CITA, University of Toronto, 60 St George Str,
Toronto, ON M5S 1A1, Canada\\ 
    E-mail: \email{kofman@cita.utoronto.ca}} 
\author{Andrei Linde\thanks{On leave of 
absence from Stanford University until 1 September 2000}\\  
    Theory Division, CERN CH 1211 Geneva 23, Switzerland\\ 
    E-mail: \email{linde@physics.stanford.edu}\\
    \email{http://physics.stanford.edu/linde}} 
\author{Igor Tkachev \\  
    Institute for Theoretical Physics, ETH,
H\"{o}nggerberg, CH-8093, Zurich, Switzerland\\ 
    E-mail: \email{Igor.Tkachev@cern.ch}}  
\received{\today}        
 \preprint{CERN-TH/2000-069\\ CITA-2000-06\\ \hepph{0004024}\\ February 29,
2000}

\abstract{Preheating after inflation may lead to nonthermal phase 
transitions with symmetry restoration. These phase transitions may 
 occur even if the total energy density of fluctuations produced during 
reheating is relatively small as compared with the vacuum energy in the 
state with restored symmetry. As a result, in some inflationary models 
one encounters a secondary, {\it nonthermal} stage of 
 inflation due to symmetry 
restoration after preheating. We review the theory of nonthermal phase 
transitions and make a prediction about the expansion factor during the 
secondary inflationary stage. We then present the results of lattice 
simulations which verify these predictions, and discuss possible 
implications of our results for the theory of formation of topological 
defects during nonthermal phase transitions. }  
 
 
\begin{document}

\section{Introduction}

The theory of cosmological phase transitions is usually associated with 
 symmetry restoration due to high temperature effects and 
the subsequent symmetry breaking which occurs as the temperature 
decreases in an expanding universe 
\cite{Kirzhnits,Weinb,Kirzhnits2,book}. A particularly important version 
of this theory is the theory of first order cosmological phase 
transitions developed in \cite{Kirzhnits2}. It served as a basis for the 
first versions of inflationary cosmology \cite{infl}, as well as for the 
theory of electroweak baryogenesis \cite{Shap}. 

Recently it was pointed out that preheating after inflation \cite{KLS} 
may rapidly produce a large number of particles that for a long time 
remain in a state out of thermal equilibrium. These particles may lead to 
specific {\em nonthermal} cosmological phase transitions 
\cite{ptr1,ptr2}. In some cases these phase transitions are first order 
\cite{bubble,Copeland}; they occur by the formation of bubbles of the 
phase with spontaneously broken symmetry inside the metastable symmetric 
phase. If the lifetime of the metastable state is large enough for the 
energy density of fluctuations to be diluted, one may encounter a short 
secondary stage of inflation {\it after} preheating \cite{ptr1}. Such a 
secondary inflation stage, if it occurs late enough, could be important 
in solving the moduli and gravitino problems.  In this respect  secondary 
``nonthermal''  inflation due to preheating may be an alternative to the 
``thermal inflation'' \cite{LS}, suggested for solving these problems.  

In this paper we will briefly present the theory of such phase 
transitions and then give the results of numerical lattice simulations 
that directly demonstrated the possibility of such brief inflation. We 
will also discuss possible implications of our results for the theory of 
formation of topological defects during nonthermal phase transitions. A 
detailed description of numerical methods used in our work will be given 
in   the appendix. 

\section{Theory of the phase transition}
Consider a set of scalar fields with the potential 
\begin{equation}\label{model}
V(\phi,\chi)= {\lambda \over 4}(\phi^{2}-v^{2})^{2}+ {g^{2}\over 2} 
\phi^{2} \chi^{2} \; . \label{p1} 
\end{equation}
The inflaton  field $\phi$ has a double-well potential and interacts with 
an $N$-component scalar field $\chi$;\, $\chi^2 \equiv \sum_{i=1}^N 
\chi_i^2$. For simplicity, the field $\chi$ is taken to be massless and 
without self-interaction. The fields couple minimally to gravity in a FRW 
universe with a scale factor $a(t)$. 

The potential $V(\phi,\chi)$ has minima at $\phi = \pm v$, $\chi = 0$ and 
a local maximum  in the $\phi$ direction at $\phi = \chi= 0$ with 
curvature $V_{,\phi\phi} = - \lambda v^2$. The effective potential 
acquires corrections due to quantum and/or thermal fluctuations of the 
scalar fields \cite{Kirzhnits,Kirzhnits2,book}, 
\begin{equation}\label{p4ax}
\Delta V = {3\over 2} \lambda \langle \phi^2 \rangle \phi^2 + {g^2\over 
2} \langle \chi^2\rangle \phi^2 + {g^2\over 2}\langle \phi ^2\rangle 
\chi^2 +..., 
\end{equation} 
 where we have written only the 
leading terms depending on $\phi$ and $\chi$. The effective mass squared 
of the field $\phi$ is given by 
\begin{equation}\label{p4}
m_{\phi}^2 = -m^2 + 3 \lambda \phi^2 + 3\lambda \langle 
 \phi ^2\rangle + g^2\langle \chi^2\rangle,
\end{equation}
  where $m^2 = \lambda v^2$. Symmetry is restored, i.e. $\phi =0$ becomes a 
stable equilibrium point, when the fluctuations $\langle \phi^2\rangle, 
\langle \chi^2\rangle$ become sufficiently large to make the effective 
mass squared positive at $\phi=0$.

For example, one may consider matter in thermal equilibrium. Then, in the 
large temperature limit, one has 
$
\langle  \phi ^2\rangle = \langle  \chi_i ^2\rangle = {T^2\over 12}. 
$
The effective mass squared of the field $\phi$ 
\begin{equation}\label{p4a}
m_{\phi ,eff}^2 = -m^2 + 3 \lambda \phi^2 + 3\lambda \langle 
 \delta\phi^2\rangle +  g^2\langle \chi^2\rangle
\end{equation}
is positive and symmetry  is restored  (i.e. $\phi =0$ is the stable 
equilibrium point) for $T > T_c$, where $T^2_c = {12 m^2\over 3\lambda + 
N g^2} \gg m^2$. At this temperature the energy density of the gas of 
ultrarelativistic particles is given by 
\begin{equation}\label{p5} 
\rho = {\cal N}(T_c) {\pi^2\over 30} T_c^4 = {24\, m^4 {\cal 
N}(T_c)\pi^2\over 5\, (3\lambda + Ng^2)^2} \ {}. 
\end{equation} 
Here ${\cal N}(T)$ is the effective number of degrees of freedom at large 
temperature, which in realistic situations may vary from $10^2$ to 
$10^3$. We will assume that $N g^2 \gg \lambda$, see below. For   $g^4 
< {96 {\cal N}(T_c)\pi^2\over 5N^2} \lambda$ the thermal energy at 
the moment of the phase transition  is greater than the vacuum energy 
density $V(0) = {m^4\over 4\lambda}$, which means that the phase 
transition does not involve a stage of inflation. 

In fact, the phase transition with symmetry breaking occurs not at $T> 
T_c$, but somewhat earlier \cite{Kirzhnits2}. To understand this effect 
let us compare the temperature $T_c \sim m/(\sqrt N g)$ and the mass 
$m_\chi = g\phi$ of the $\chi$ particles in the minimum of the 
zero-temperature effective potential at $\phi = v  = m/\sqrt \lambda$. 
One can easily see that $m_\chi \gg T_c$ for $N g^4 \gg \lambda$. This 
means that for  $N g^4 \gg \lambda$ the temperature $T_c$ is insufficient 
to excite perturbations of the fields $\chi_i$ at  $\phi = v$. As a 
result, these perturbations do not change the shape of the effective 
potential $\phi = v$. Thus the potential  at $T$ slightly above $T_c$  
has its old zero-temperature  minimum at   $\phi = v$, as well as the 
temperature-induced minimum at $\phi = 0$. Symmetry breaking occurs as a 
first-order phase transition due to formation of bubbles of the phase 
with    $\phi \approx v$ at some temperature above $T_c$ when the minimum 
at   $\phi = v$ becomes deeper than the minimum at $\phi = 0$, and the 
probability of bubble formation becomes sufficiently large. A more 
detailed investigation in the case $N = 1$ shows that the phase 
transition is first order under a weaker condition $g^3 \gg \lambda$ 
\cite{Kirzhnits2}.

In the case $N g^4 
> 10^2 \lambda$ the phase transition occurs after a
 secondary
 stage of inflation.
 In this regime radiative corrections 
 are important. They lead to the creation of 
a local minimum of $V(\phi,\chi)$ at $\phi = 0$ even at zero temperature, 
and the phase transition occurs from a strongly supercooled state 
\cite{Kirzhnits2}. That is why the first models of inflation required 
supercooling at the moment of the phase transition \cite{infl}. 

In supersymmetric theories  one may have $N g^4 \gg 10^2 \lambda$ and 
still have a  potential which is flat near the origin due to cancellation 
of quantum corrections of bosons and fermions \cite{LS}. In such cases 
the thermal energy becomes smaller than the vacuum energy at $T < T_0$, 
where $T^4_0 = {15 \over 2 {\cal N} \pi^2}m^2 v^2$. Then one may   have a 
short stage of inflation which begins at $T \sim T_0$ and ends at $T = 
T_c$. During this time the universe may inflate by the factor 
\begin{equation}\label{p5a}
{a_c\over a_0} = {T_0\over T_c} \sim 10^{-1} \Bigr({g^4\over 
\lambda}\Bigl)^{1/4}  . 
\end{equation}

Similar phase transitions may occur much more efficiently  prior to  
thermalization, due to the anomalously large fluctuations $\langle 
\phi^2\rangle$ and $\langle\chi^2\rangle$ produced during preheating 
\cite{ptr1,ptr2}. These fluctuations can change the shape of the 
effective potential and lead to symmetry restoration. Afterwards, the 
universe expands, the values of  $\langle \phi^2\rangle$ and 
$\langle\chi^2\rangle$ drop down, and the phase transition with symmetry 
breaking occurs. 

An interesting feature of nonthermal phase transitions is that they may 
occur even in theories where the usual thermal phase transitions do not 
happen. The main reason can be understood as follows. Suppose reheating 
occurs due to the decay of a scalar field with energy density $\rho$. If 
this energy is instantly thermalized, then one obtains relativistic 
particles with energy density $O(T^4)$ which in the first approximation 
can be represented as 
 $\rho \approx E^2(\langle\phi^2\rangle + \langle\chi^2\rangle)$. Here
$E \sim T \sim \rho^{1/4}$ is a typical energy of a particle in thermal 
equilibrium. After preheating, however, one has particles $\phi$ and 
$\chi$ with much smaller energy but large occupation numbers. As a 
result, the same energy release may create much greater values of 
$\langle\phi^2\rangle$ and $\langle\chi^2\rangle$ than in the case of 
instant thermalization. This may lead to symmetry restoration after 
preheating even if the symmetry breaking occurs on the GUT scale, $v \sim 
10^{16}$ GeV \cite{ptr1,ptr2}. 

The main conclusions of \cite{ptr1,ptr2} have been confirmed by detailed 
investigation using lattice simulations in 
\cite{bubble,strings,kawasaki,Copeland}. One of the main results obtained 
in \cite{bubble} was that for sufficiently large $g^2$ nonthermal phase 
transitions are first order. They occur from a  metastable vacuum at 
$\phi = 0$ due to the creation of bubbles  with $\phi \not = 0$. This 
result is very similar to the analogous result in the theory of thermal 
phase transitions \cite{Kirzhnits2}.  According to \cite{bubble}, the   
necessary conditions for this transition to occur and to be of the first 
order can be formulated as follows: 

(i) At the time of the phase transition, the point $\phi=0$ should be a 
local minimum of the effective potential. From (\ref{p4}), we see that 
this means that $Ng^2 \langle \chi_i^2 \rangle > \lambda v^2$. 

(ii) At the same time, the typical momentum $p_*$ of $\chi_i$ particles 
should be smaller than $gv$. This is the condition of the existence of a 
potential barrier. Particles with momenta $p<gv$ cannot penetrate the 
state with $|\phi|\approx v$, so they cannot change the shape of the 
effective potential at $|\phi| \approx v$. Therefore, if both conditions 
(i) and (ii) are satisfied, the effective potential has a local minimum 
at $\phi=0$ and two degenerate  minima at $\phi \approx \pm v$. 

(iii) Before the minima at $\phi\approx \pm  v$ become deeper than the 
minimum at $\phi=0$, the inflaton's zero mode should decay significantly, 
so that it performs small oscillations near $\phi=0$. Then, after the 
minimum at $|\phi| \approx v$ becomes deeper than the minimum at $\phi = 
0$, fluctuations of $\phi$ drive the system over the potential barrier, 
creating an expanding bubble.

The investigation performed in \cite{bubble} confirmed that for 
sufficiently large  $g^2$ and $N$ these conditions are indeed satisfied 
and the phase transition is first order. One may wonder whether for $g^2 
\gg \lambda$ one may have a stage of inflation in the metastable vacuum 
$\phi = 0$. 

Analytical estimates of Ref. \cite{ptr1} suggested that this is indeed 
the case, and the degree of this inflation for $N = 1$ is expected to be 
\begin{equation}\label{p5b}
{a_c\over a_0}   \sim  \Bigr({g^2\over \lambda}\Bigl)^{1/4} , 
\end{equation}
which is much greater than the number  $10^{-1} \Bigr({g^4\over 
\lambda}\Bigl)^{1/4}$ in the thermal inflation scenario. One could also 
expect that the duration of inflation, just like the strength of the 
phase transition, increases if one considers N fields $\chi_i$ with $N 
\gg 1$. 

However, the theory of preheating is extremely complicated, and there are 
some factors which  could not be adequately taken into account in the 
simple estimates of \cite{ptr1}. The most important factor is the effect 
of rescattering of particles produced during preheating \cite{lattice}. 
This effect  
tends to shut down  the resonant production of particles and thus shorten 
or prevent entirely the occurrence of a secondary stage of inflation. 
Thus the estimates above reflect the maximum degree of inflation possible 
for a given set of parameter values, but in practice the expansion factor 
will be somewhat smaller than these predictions. The only way to fully 
account for all the effects of backreaction and expansion is through 
numerical lattice simulations. In our paper we used a generalized version 
of the method of lattice simulations developed in \cite{lattice,bubble}. 
In the next section we will describe the basic features of our method and 
describe our main results.  A detailed description of  the lattice 
simulations will be given in the  appendix. 

\section{Simulation Results and Their Interpretation}\label{simulations}

In our paper we will take  $\lambda \approx 10^{-13}$, which gives the 
proper magnitude of inflationary perturbations of density 
\cite{book,kolb}. We assume that $g^2 \gg \lambda$, and consider $v 
\approx 10^{16}$ GeV, which corresponds to the GUT scale. A numerical 
investigation of preheating in 
 the model (\ref{model})
   was first performed in 
\cite{bubble}. 
  The authors found a strongly first order phase 
transition. The strength of the phase transition increased with an 
increase of $g^2/\lambda$ and of the number $N$ of the fields $\chi_i$. 
However, for the parameters of the model studied in \cite{bubble} 
($g^2/\lambda \approx 200$) there was no inflation during symmetry 
restoration. This is not unexpected because the estimates discussed above 
indicated that the expansion of the universe during the short stage of 
nonthermal inflation cannot be greater than $\bigl({g^2\over \lambda 
}\bigr)^{1/4}$. 

Keeping in mind that $\lambda$ in this model is extremely small, one 
would expect that in realistic versions of this model one may have  
$g^2/\lambda$ as large as $10^{10}$, which could lead to a relatively 
long stage of inflation. However, for very large $g^2/\lambda$ our 
analytical estimates are unreliable, and lattice simulations become 
extremely difficult: One needs to have enormously large lattices to keep 
both infrared and ultraviolet effects under control. 

\FIGURE{\epsfig{file=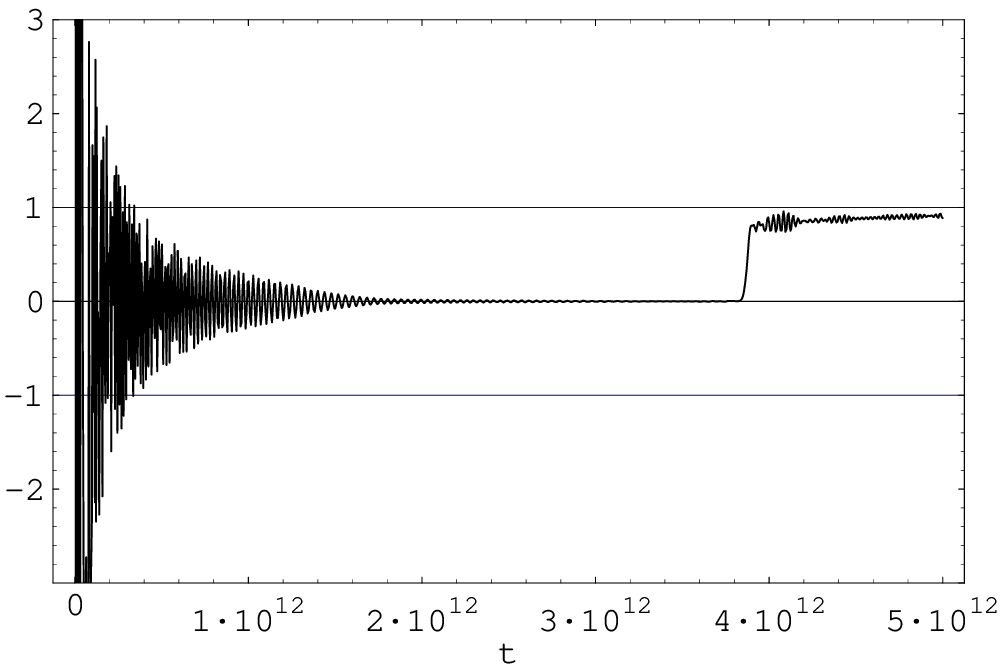,width=13cm} 
        \caption{The spatial average of the inflation field $\phi$ as a
function of time. The field $\phi$ is shown in units of $v$, the symmetry 
breaking parameter. Time is shown in Planck units.}  
    \label{phiplot}} 

To mimic the effects of large $g^2$, we considered a large number of the 
fields $\chi_i$. We have performed simulations for $g^2/\lambda = 800$ 
and $N = 19$. With these parameters the strength of the phase transition 
became much greater, and there was a short stage of inflation prior to 
the phase transition. The details of our calculation and an explanation 
of our methods are given in the  appendix. Here we only present the main 
results.

The simulation showed that the oscillations of the inflaton field 
decreased until the field was trapped near zero. It remained there until 
the moment of the phase transition when it rapidly jumped to its symmetry 
breaking value, as shown in Fig.~\ref{phiplot}. 

The trapping of the field occurred because of the corrections to the 
effective potential induced by the particles $\phi$ and $\chi$ produced 
during preheating, just like in the theory of high-temperature phase 
transitions. In our case, however, this effect has some unusual features.

To first order in $g^2$, the 
 leading contribution to the equation of motion $\ddot\phi = - V'$ is given
by 
 $g^2\phi \langle \chi^2 \rangle$, where
\begin{equation}
 \langle \chi^2 \rangle  \approx { N  
 \over {2\pi^2 }}~ \int\limits_0^{\infty} {{n_k\, k^2\, dk}\over
\omega_k(\phi)}  \ . \label{100} 
\end{equation}
Here  $\omega_k =\sqrt{k^2 +g^2 (\phi^2+\langle \phi^2 \rangle)}$ is the 
energy of  $\chi_i$ particles with momentum $k$ and $n_k$ is their 
occupation number; $\phi$ is the homogeneous component of the field. For 
$\phi\ll \sqrt{\langle \phi^2 \rangle}$, one has  
\begin{equation}
 \langle \chi^2 \rangle_{_{\phi = 0}}  \approx { N  
 \over {2\pi^2 }}~ \int\limits_0^{\infty} 
 {n_k\, k^2\, dk\over \sqrt{k^2 +g^2 \langle \phi^2 \rangle}}  \ .
\label{100x} 
\end{equation}
This quantity does not depend on $\phi$; it can be evaluated using our 
lattice simulations when the field $\phi$ oscillates near $\phi = 0$. It 
leads to the usual quadratic correction to the effective potential, see 
Eq. (\ref{p4ax}). This correction adequately describes the change of the 
shape of the effective potential for $\phi$ smaller than the amplitude of 
the oscillations of this field, because most of the time prior to the 
moment of the phase transition this amplitude is much smaller than 
$\sqrt{\langle \phi^2 \rangle}$. 

However, if we want to evaluate the effective potential at all values of 
$|\phi|$ from $0$ to $v$, rather than for $\phi$ similar to the amplitude 
of the oscillations, then one should take into account that for 
sufficiently large $|\phi|$ the term $g|\phi|$ becomes greater than 
$g\sqrt{\langle \phi^2 \rangle}$ and than the typical momentum $k$ of 
particles  $\chi_i$.  In this case the main contribution to $\langle 
\chi^2 \rangle$ is given by nonrelativistic particles with  $\omega_k 
\approx +g|\phi|$, and one has 
\begin{equation}
\langle \chi^2 \rangle  \approx { N  
 \over {2\pi^2 }}~ \int\limits_0^{\infty} {{n_k\, k^2\, dk}\over
g|\phi|} = {n_\chi \over g|\phi|}\ ,\ \label{100axx} 
\end{equation}
where $n_\chi$ is the total density of all types of $\chi_i$ particles. 
This implies that at large $|\phi|$ the effective potential acquires a 
correction 
\begin{equation}
\delta V \approx    g|\phi| n_\chi. 
\end{equation} 
Thus, instead of being quadratic or cubic in $|\phi|$, as one could 
expect from the analogy with the high-temperature theory 
\cite{Kirzhnits2,Copeland}, the corrections to the effective potential 
 at large $|\phi|$  are proportional to $|\phi|$ \cite{KLS}. 
 
 The combination of these two 
 types of corrections to the effective potential (quadratic at small $|\phi|$
and linear at large 
 $|\phi|$) leads to the symmetry restoration that we have found in our
lattice 
 simulations.

It is instructive to look in a more detailed way at the small region near 
the time of the phase transition. The first of the graphs in Fig. 
\ref{near} shows the oscillations of the field $\phi$ soon before the 
phase transition, whereas the second one shows  these oscillations soon 
afterwards.

\    

\FIGURE{\epsfig{file=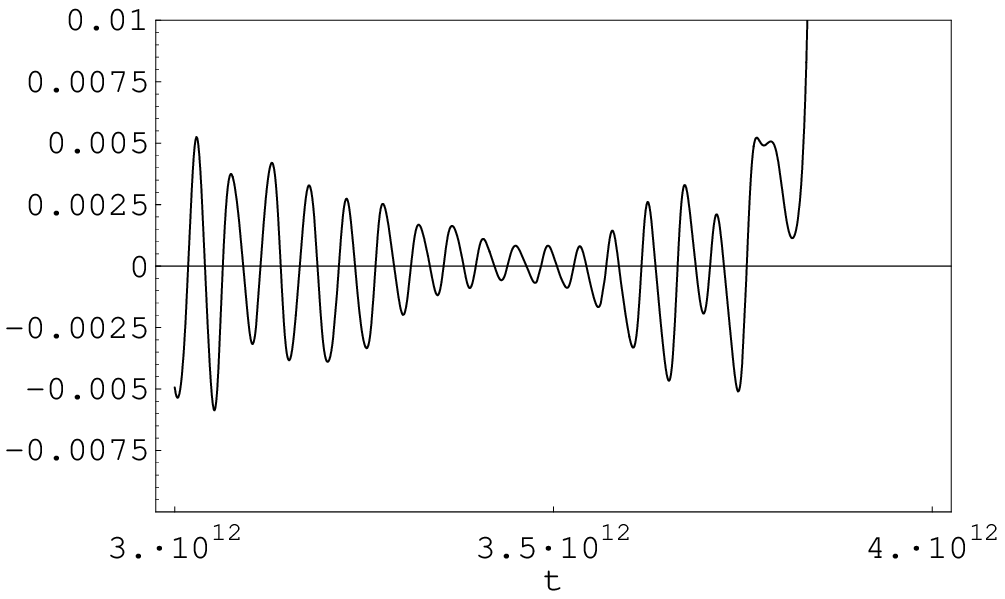,width=7.4cm} \hskip 0.5cm 
 \epsfig{file=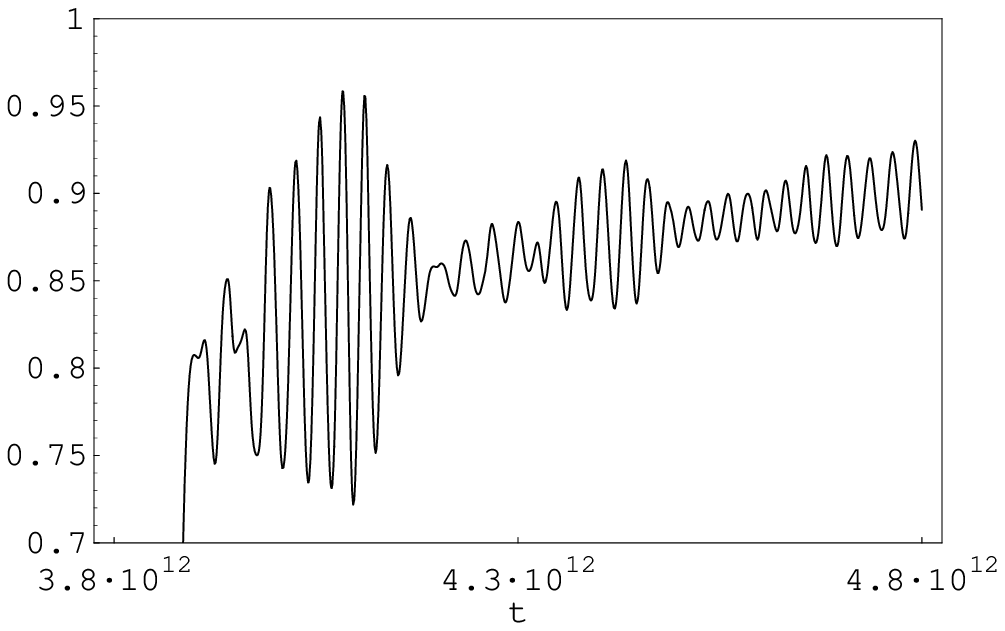, width=7cm} 
        \caption{The spatial average of the inflation field $\phi$ as a
function of time in the vicinity of the phase transition. The left figure 
shows the field just before the phase transition and at the moment of the 
transition. The field oscillates with an amplitude approaching $10^{-3} 
v$. The right figure shows the field $\phi$ after the phase transition, 
when it oscillates near the (time-dependent) position of the minimum of 
the effective potential at $\phi \approx v$.  Time is shown in Planck 
units.}  
    \label{near}}

First of all, one can see that just before the phase transition the field 
oscillates with an amplitude three orders of magnitude smaller than $v$, 
which is a clear sign of symmetry restoration. Another interesting 
feature is that the frequency of oscillations  does not vanish  as we 
approach the phase transition, but remains  nearly constant. Moreover, 
this frequency is only about two times smaller than the frequency of 
oscillations after the phase transition, which is equal to $\sqrt 2 m$. 
Note that the frequency of the oscillations is determined by the 
effective mass of the scalar field, which is given by  the curvature of 
the effective potential:   $m^2_\phi = V''$. This means that  at the 
moment of the phase transition the effective potential has a deep minimum 
at $\phi = 0$ with curvature $V'' \sim + m^2$, i.e. the phase transition 
is strongly first order. Such phase transitions should occur due to the 
formation of bubbles containing nonvanishing field $\phi$.

Indeed, we have found that this transition occurred in a nearly spherical 
region of the lattice that quickly grew to encompass the entire space. 
The growth of this region of the new phase is shown in Fig. 
\ref{bubbleplot}. The nearly perfect sphericity of this region is an 
additional  indication that the transition was strongly first-order. In 
comparison, the bubble observed in the lattice simulations of 
\cite{bubble} for  $g^2/\lambda \approx 200$ was not exactly spherically 
symmetric.

\FIGURE{\epsfig{file=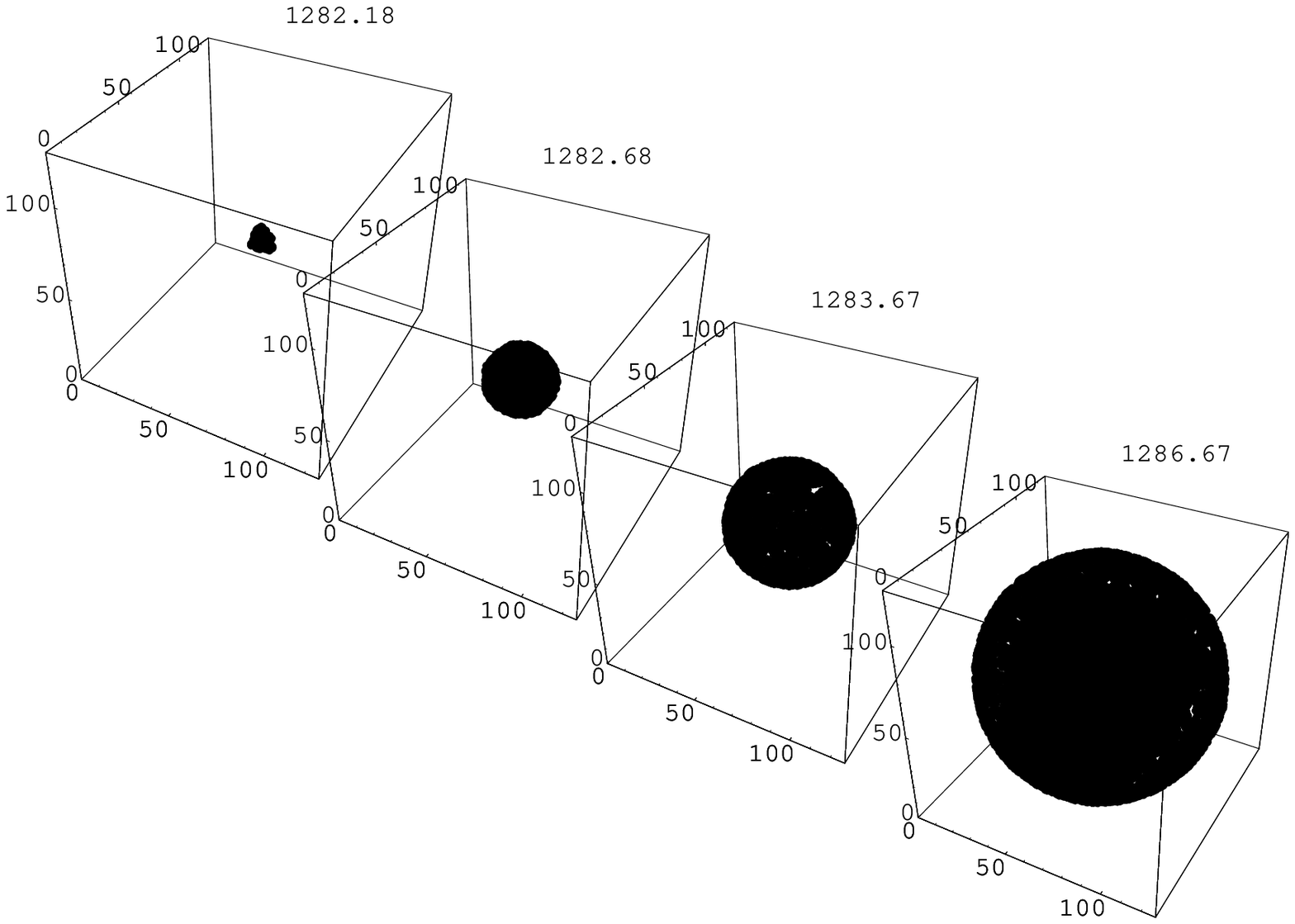,width=13cm}
\caption{These plots show the region of space in which symmetry
breaking has occurred at four successive times labeled by the value of
the conformal time $t_{pr}$.} \label{bubbleplot}}

The first order phase transition and bubble formation seen in our 
simulations can be understood as a result of gradual accumulation of 
classical fluctuations $\delta \phi(t, \vec x)$. These fluctuations 
stochastically climb up from $\phi=0$ towards the local maximum $\phi_*$ 
of the effective potential. Consider the regions in which $\delta \phi(t, 
\vec x) > \phi_*$.  If the probability of formation of such regions is 
small because they correspond to high peaks of the random field $\phi$, 
then these regions will have a nearly spherical shape and can be 
represented by spherical surfaces of radius $R_*$ (bubbles). If the 
radius $R_*$ is small, gradient terms will prevent the field $\phi$ 
inside the region from rolling down towards the global minimum at 
$\phi=v$ (subcritical bubble).  If $R_*$ is large enough, the gradient 
terms cannot push the field back to the metastable state $\phi = 0$ and 
the field inside the bubble rolls towards the global minimum, forming a 
bubble of ever increasing radius. This process can be described within 
the stochastic approach to tunneling proposed in \cite{AL92}. Typically, 
the gradient terms cannot win over the potential energy terms if $R_* > 
O(|m_\phi^{-1}|)$, where $m_\phi$ corresponds to the effective mass of 
the scalar field in the interior of the bubble. This provides an estimate 
for the initial size of the bubble $R_* \sim O(|m_\phi^{-1}|)$. 

The phase transition occurs from a state with energy density dominated by 
the vacuum energy density $V(0)$. Figure \ref{aplot} shows the scale 
factor $a$ as a function of time. The curvature becomes slightly positive 
at the time before the phase transition, which indicates a short stage of 
exponential growth of the universe.  Because the curvature is hard to see 
in figure \ref{aplot} we have also plotted the second derivative 
$\ddot{a}$ in figure \ref{a2dplot}. While the inflaton is trapped in the 
false vacuum state, $\ddot a$ becomes positive, indicating a brief stage 
of inflation. 

\FIGURE{\epsfig{file=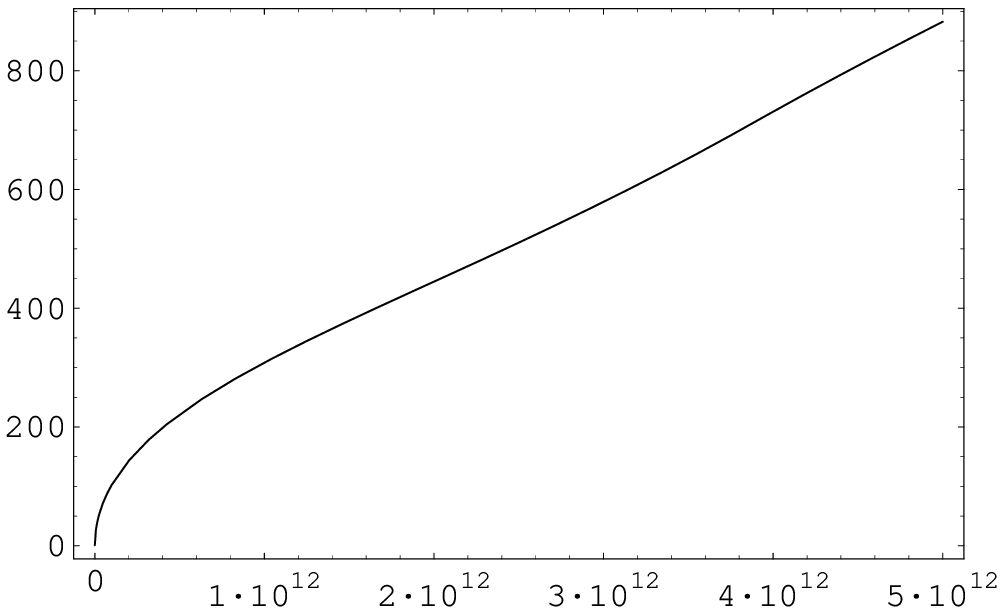,width=13cm} 
        \caption{The scale factor $a$ as a function of time. 
        In the beginning $a \sim \sqrt t$, which is a 
        curve with negative curvature, but then at some stage
        it begins to turn upwards, 
       indicating a short stage of inflation.} \label{aplot}}

\FIGURE{\epsfig{file=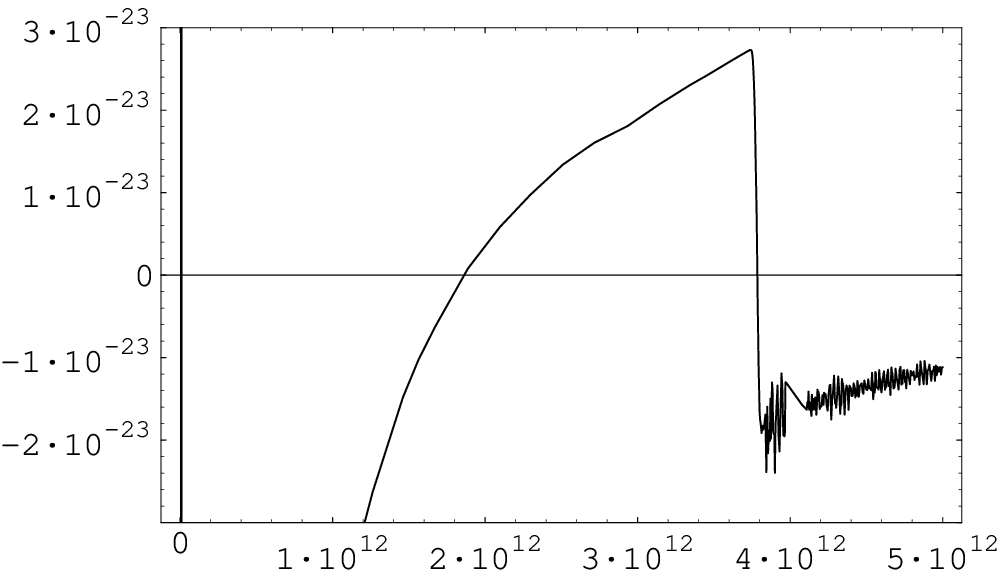,width=13cm} \caption{The second 
        derivative of the scale factor, $\ddot a$.  A universe
        dominated by ordinary matter (relativistic or nonrelativistic)
        will always have $\ddot{a}<0$, whereas in an inflationary
        universe $\ddot{a}>0$. We see that starting from the moment $t
        \sim 2\times 10^{12}$ (in Planck units) the universe
        experiences accelerated (inflationary) expansion.}
        \label{a2dplot}}

 Another signature of inflation is 
an equation of state with negative pressure. Figure \ref{pressureplot} 
shows the parameter $\alpha=p/\rho$, which becomes negative during the 
metastable phase. At the moment of the phase transition the universe 
becomes matter dominated and the pressure jumps to nearly $0$. 

\FIGURE{\epsfig{file=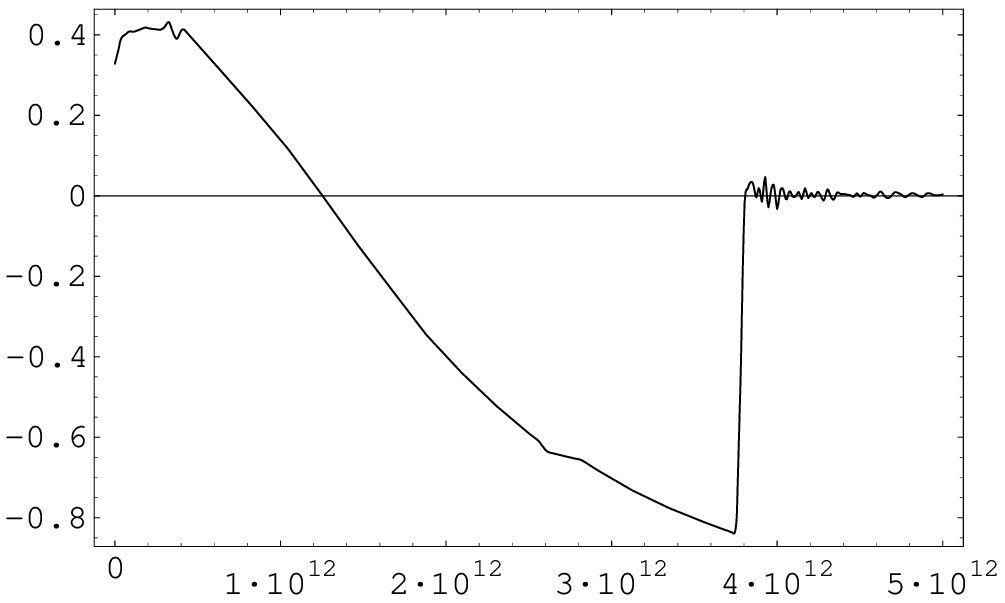,width=13cm} 
        \caption{The ratio of pressure to
energy density $p/\rho$. (Values were time averaged over short time 
scales to make the plot smoother and more readable.)} 
\label{pressureplot}}

From the beginning of this inflationary stage (roughly when the pressure 
becomes negative) to the moment of the phase transition the total 
expansion factor is 2.1. As expected this is of the same order but 
somewhat lower than the predicted maximum, $\left({g^2 \over 
\lambda}\right)^{1/4} \approx 5.3$. We can thus conclude that it is 
possible to achieve inflation for parameters for which this would not 
have been possible in thermal equilibrium ($g^4 \ll \lambda$). In our 
simulation we have shown the occurrence of a very brief stage of 
inflation. This stage may be much longer for larger (realistic) values of 
$g^2/\lambda$. However, to check whether this is indeed the case one 
would need to perform a more detailed investigation on a lattice of a 
much greater size. 

\

\section{Nonthermal phase transitions and production of topological defects}

In this section we would like to discuss possible implications of our 
investigation for the theory of production of topological defects after 
preheating \cite{bubble,strings,kawasaki}. 

 The bubbles that appear after the phase transition  can 
contain either positive or negative field, $\phi = \pm v$. If bubbles of 
either type are formed with comparable probability, then after the phase 
transition the universe becomes divided into  nearly equal numbers of 
domains with $\phi = \pm v$, separated by domain walls. Such domain walls 
would lead to disastrous cosmological consequences, which would rule out 
the models where this may happen \cite{ptr1,ptr2}. 

In general, the number of bubbles with  $\phi = + v$ may be much greater 
(or much smaller) than the number of bubbles with  $\phi = - v$.  Then 
the domain wall problem does not appear because the bubbles with $\phi = 
+ v$ would rapidly eat all their competitors with $\phi = - v$ (or {\it 
vice versa}). This may happen, for example, if the moment of the bubble 
production is determined  by the coherently oscillating 
 scalar field $\phi$. In such a case, 
after oscillating a bit near the top of the effective potential, the 
field $\phi$ may wind up in the same minimum of the effective potential 
everywhere in the universe.  

To investigate the domain wall problem in our model one would need to 
repeat the calculation many times with slightly different initial 
conditions or to make them in a box of a much greater size that would 
allow one to see many bubbles simultaneously.    Fortunately, the results 
obtained in our study may be sufficient to give an answer to this 
question without extremely large simulations. 

First of all, according to the stochastic approach \cite{AL92} to the 
theory of tunneling with bubble formation \cite{coleman,AL83}, the 
bubbles of the field $\phi$ are created as a result of the accumulation 
of long-wavelength fluctuations of the scalar field with momenta $k$ 
smaller than the typical mass scale $m_\phi$ associated with this field, 
see Section \ref{simulations}. In our case this mass scale is related to 
the frequency of oscillations of the scalar field at the moment of the 
phase transition. At that moment the leading contribution to the 
fluctuations $\langle \phi^2 \rangle$ is given by fluctuations with 
momenta much smaller than $m_\phi$. We calculated the value of the 
long-wavelength component of $\sqrt{\langle \phi^2 \rangle}$ and found 
that it is (approximately) of the same order as the amplitude of 
oscillations of the field $\phi$ at the moment of the phase transition.  
The existence of a first-order phase transition suggests that the 
probability of bubble formation must have been exponentially suppressed 
during the metastable stage. Such suppression would only occur only if 
the amplitude of fluctuations required to form a bubble of the new phase 
was much larger than $\sqrt{\langle \phi^2 \rangle}$ \cite{AL92}, which 
would in turn mean the required amplitude was much greater than the 
amplitude of oscillations of $\phi$. This suggests that the probability 
of the bubble formation is almost entirely determined by the incoherent 
fluctuations of the field $\phi$ rather than by the small coherent 
oscillations of this field. Consequently, the probability of formation of 
bubbles containing $\phi = + v$ in the first approximation must be equal 
to the probability of formation of bubbles containing $\phi = - v$. 

To make this statement more reliable  one would need to estimate the 
amplitude of the long-wavelength fluctuations of the field $\phi$ in a 
more precise way, which would involve using lattices of a greater size. 
However, there is additional evidence suggesting that the number of 
bubbles with positive and negative $\phi$ must be approximately equal to 
each other. 

Indeed, as we have seen, the curvature of the effective potential 
remained approximately constant during dozens of oscillations of the 
field $\phi$ prior to the moment of the phase transition. This suggests 
that the shape of the effective potential and, consequently, the 
probability of the tunneling, did not change much during a single 
oscillation. Therefore one may expect that, within a single oscillation, 
the probability of a bubble forming when the oscillating field $\phi$ was 
negative was approximately the same as the probability of the bubble 
forming when it was positive.

If the number of bubbles with positive and negative $\phi$ is 
approximately equal to each other, the phase transition leads to the 
formation of dangerous domain walls, which rules out our model 
\cite{zeld}. If correct, this is a rather important conclusion which 
shows that the investigation of nonthermal phase transition may rule out 
certain classes of inflationary models which otherwise would seem quite 
legitimate \cite{ptr1,ptr2}. 

But this conclusion does not imply that {\it all} theories where the 
nonthermal phase transition is strongly first order are ruled out. For 
example, one may consider a model (\ref{model}) with $\phi$ being not a 
real but a complex field, $\phi = {1\over \sqrt 2} (\phi_1 + i \phi_2)$, 
$\vert \phi\vert^2 = {1\over 2} (\phi_1^2 + \phi_2^2)$. Since the main 
contribution to the effective potential of the field $\phi$ in the theory 
(\ref{model}) is given not by the field(s) $\phi$ but by the fields 
$\chi_i$, we expect that this generalization will not lead to a 
qualitative modification of our results. In particular, we expect that 
for sufficiently large $N$ and $g^2/\lambda$ the phase transition will be 
strongly first order and there will be a short stage of inflation after 
preheating. However, in the new model we will have strings instead of 
domain walls. 

A similar model  in the absence of interaction of the fields $\phi$ with 
the fields $\chi$ was studied in  \cite{strings,kawasaki}. It was argued 
that even in this case infinite strings may be formed. The theory of 
galaxy formation due to cosmic strings is currently out of favor, but it 
is certainly true that cosmic strings produced after inflation  may add 
new interesting features to the standard theory of  formation of the 
large-scale structure of the universe \cite{LR}

The possibility of strongly first order phase transitions induced by 
preheating in models with $g^2 \gg \lambda$ adds new evidence that 
infinite strings can  be produced after nonthermal phase transitions. 
Indeed, infinite strings may not be produced if the direction in which 
the field $\phi$ falls from the point $\phi = 0$ at the moment of the 
phase transition  is determined by the oscillations of the field $\phi$. 
If, just as in the case discussed above, the amplitude of these 
oscillations are much smaller than $\sqrt{\langle \phi^2 \rangle}$ at the 
moment of the phase transition, then infinite strings are indeed formed.

\section{Conclusions}

The results of our lattice simulation confirm our expectations that 
preheating may lead to nonthermal phase transitions even in those 
theories where spontaneous symmetry breaking occurs at the GUT scale, $v 
\sim 10^{16}$ GeV. Some time ago this question was intensely debated in 
the literature. Some authors claimed that nonthermal phase transitions 
induced by preheating are impossible, and the notion of the effective 
potential after preheating is useless. In our opinion, Figs. 1, 2 and 3 
give a clear answer to this question. In particular, Fig. 1 shows that 
90\% of the time from the end of inflation to the moment of symmetry 
breaking the field $\phi$ oscillates about $\phi = 0$ with an amplitude 
much smaller than $v$. This could happen only because the corrections to 
the effective potential induced by particles $\phi$ and $\chi$ change the 
shape of $V(\phi)$ near $\phi = 0$, turning its maximum into a deep local 
minimum.

In some theories, this effect may lead to production of superheavy 
strings, which may have important cosmological implications for the 
theory of formation of the large scale structure of the universe. In some 
other theories, these phase transitions may lead to excessive production 
of monopoles and domain walls. This may rule out a broad class of  
otherwise acceptable inflationary models. 

In this paper we have shown that under certain conditions  a nonthermal 
phase transition may lead to a short secondary stage of inflation. It 
would be interesting to study the possibility that a secondary stage of 
inflation induced by preheating could  help solve the moduli and 
gravitino problems.  The answer to this question will be strongly 
model-dependent because gravitinos can be produced by the oscillating 
scalar field even after the secondary inflation \cite{KKLP,GTR}. 
Independently of all practical implications, the possibility of a 
secondary stage of inflation induced by preheating seems very interesting 
because it clearly demonstrates the potential importance of 
nonperturbative effects in post-inflationary cosmology.

The authors are grateful to Julian Borrill for very useful discussions. 
This work was supported  by  NSERC and  CIAR and by NSF grant 
AST95-29-225. The work of G.F. and A.L. was also supported   by NSF grant 
PHY-9870115, and the work of L.K. and A.L. by NATO Linkage Grant 975389. 

\ 

\section{Appendix: The Numerical Calculations}

Our lattice program solves the classical equations of motion for a set of 
scalar fields in a Friedman-Robertson-Walker universe. These fields 
include an inflaton $\phi$ coupled to a set of matter fields $\chi_i$. 
The scale factor $a$ is also solved for self-consistently. In this 
appendix we describe the exact equations being solved and the method used 
to solve them. 

Unless otherwise specified, all variables names refer to their bare 
physical values measured in Planck units. The quantities as they appear 
in the program have been rescaled in several ways, and these values will 
be indicated with {\it pr}, as in $\phi_{pr}$. The relations between the 
rescaled variables and their bare values are given below. Also, when we 
want to refer to a general property of the fields $\phi$ and $\chi_i$ we 
will use $\sigma$ to   indicate a generic scalar field. 

The simulations we discuss here used a grid of $128^3$ points. The 
spacing of the points was chosen so as to ensure that the ultraviolet 
cutoff imposed by the grid was higher than all physically relevant 
momenta in the problem. To check this we monitored the power spectra 
$\vert \phi_k\vert^2$ and $\vert \chi_k\vert^2$ throughout the run. On a 
log plot you can see that there is a momentum above which the slope of 
the power spectrum decreases sharply (i.e. becomes more negative), and we 
set our grid spacing to ensure that this cutoff was well below the cutoff 
of the grid. A sample spectrum for $\chi$, taken from the late stages of 
preheating, is shown in figure \ref{chispectrum}. The kink that can be 
seen at $k \approx k_{max}/2$ persisted from preheating through the phase 
transition, and a similar feature can also be seen in the $\phi$ 
spectrum. 

\FIGURE{\epsfig{file=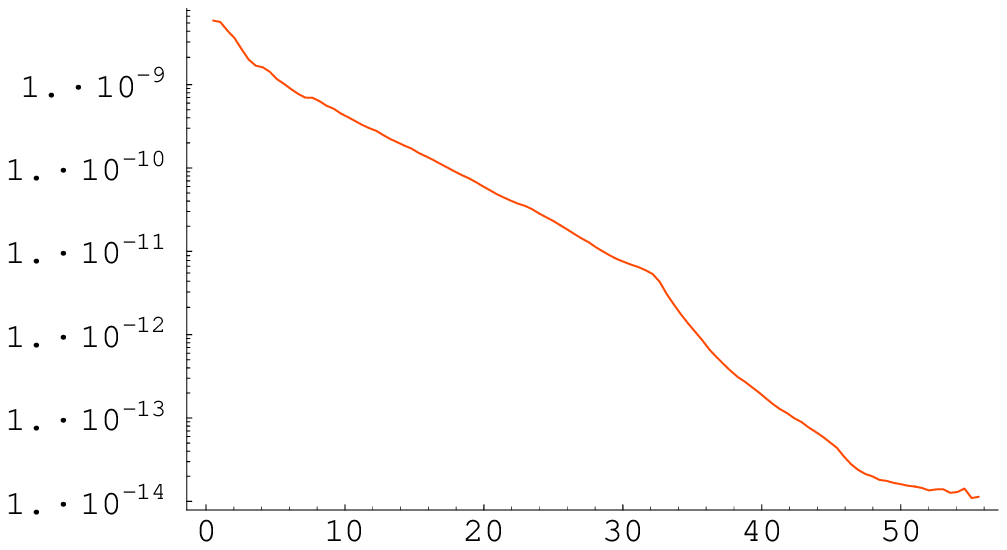,width=13cm} \caption{The 
        spectrum $\vert \chi_k\vert^2$ as a function of momentum,
        shown in the late stages of preheating. The momenta $k$ are
        shown in units of the Hubble constant at the end of
        inflation.}  \label{chispectrum}}

The grid spacing we used was 
\begin{equation}
dx = 0.133 {1 \over \sqrt{\lambda} \phi_0} \approx 1.26 \times 10^6 
M^{-1}_{P} 
\end{equation}
measured in comoving coordinates with $a=1$ at the beginning of the 
simulation. The total box size, $L = 128 dx$, was equal to slightly more 
than $10$ Hubble radii at the end of inflation. 

As a further check that the ultraviolet grid cutoff was not affecting the 
physics we did some smaller runs (i.e. with fewer fields) using these 
same parameters and another run using a grid of $64^3$ points with twice 
as large a grid spacing. Both the $64^3$ and $128^3$ runs had the same 
total box size and they showed essentially identical behavior for the 
fields, suggesting that the $128^3$ grid has more than enough modes in 
the ultraviolet. 

\subsection{The Field Equations}

We work in a FRW universe with metric 
$g_{\mu\nu}=diag\{1,-a^2,-a^2,-a^2\}$. The equations of motion for the 
scalar fields $\phi$ and $\chi_i$ derive from the potential 
\begin{equation}\label{potential}
V = {\lambda \over 4} \left(\phi^2 - v^2\right)^2 + {1 \over 2} g^2 
\phi^2 \chi_i^2 
\end{equation}
where $v$, $\lambda$, and $g$ are constant parameters and $\chi_i^2$ is 
understood to include a summation over $i$. The results reported here 
used $v=7 \times 10^{-4}$, $\lambda=9 \times 10^{-14}$, and $g^2 = 800 
\lambda$, and took $\chi$ to be a 19-component field. This potential 
gives rise to the equations of motion 
\begin{equation}
\ddot{\phi} + 3 {\dot a \over a} \dot{\phi} - {1 \over a^2}\nabla^2 \phi 
+ \left(\lambda \left(\phi^2 - v^2\right) + g^2 \chi_i^2\right) \phi = 0 
\end{equation}
\begin{equation}
\ddot{\chi_i} + 3 {\dot a \over a} \dot{\chi_i} - {1 \over a^2}\nabla^2 
\chi_i + g^2 \phi^2 \chi_i = 0 
\end{equation}

These equations can be simplified by the following variable redefinitions 
\begin{equation}\label{rescalings}
\phi_{pr} = {a \over \phi_0} \phi; \chi_{i,pr} = {a \over \phi_0} \chi_i; 
v_{pr} = {1 \over \phi_0} v; \vec{x}_{pr} = \sqrt{\lambda} \phi_0 
{\vec{x}}; \tau_{pr} = \sqrt{\lambda} \phi_0 \int {dt \over a}; ge \equiv 
{g^2 \over \lambda} 
\end{equation}
where $\phi_0$ is the initial (bare) value of the field $\phi$ and all 
other parameters come from the original equations of motion. These then 
give the rescaled equations of motion 
\begin{equation}\label{phievolution}
\phi_{pr}'' - \nabla^2_{pr}\phi_{pr} + \left(\phi_{pr}^2 - a^2 v_{pr}^2 + 
ge\ \chi_{i,pr}^2 - {a'' \over a}\right)\phi_{pr} = 0 
\end{equation}

\begin{equation}\label{chievolution}
\chi_{i,pr}'' - \nabla^2_{pr}\chi_{i,pr} + \left(ge\ \phi_{pr}^2 - {a'' 
\over a}\right)\chi_{i,pr} = 0 
\end{equation}
where derivatives are with respect to the rescaled time and distance 
variables defined above. Note that all first derivative terms have been 
eliminated, and the dependence on the coupling constants $\lambda$ and 
$g$ is now only through the ratio $g^2/\lambda$, denoted here by $ge$. 
The value of $\lambda$ itself appears only in setting the initial 
conditions (described below). The rescalings involving $\phi_0$ do not 
affect the equations of motion. 

\subsection{The Scale Factor Equation}

Our simulations calculate a single scale factor at each time, neglecting 
metric fluctuations. The equation for the scale factor $a$ is derived 
>from the following two equations 
\begin{equation}\label{adotdot}
\ddot a = -{4 \pi \over 3} (\rho + 3 p) a 
\end{equation}
\begin{equation}\label{hubble}
\left({\dot a \over a}\right)^2 = {8 \pi \over 3} \rho 
\end{equation}
where $\rho$ and $p$ are the total energy density and pressure, 
respectively. For simplicity we will first solve these for a single 
scalar field $\sigma$. The energy density and pressure can be derived 
>from the energy momentum tensor  
\begin{equation}
T_{\mu\nu} = \sigma_{,\mu}\sigma_{,\nu} - {1 \over 2} g_{\mu\nu} 
g^{\alpha\beta} \sigma_{,\alpha}\sigma_{,\beta} + g_{\mu\nu} V(\sigma). 
\end{equation}
Assuming the fields are isotropic this equation can be solved and 
compared to the usual energy momentum tensor of matter 
\begin{equation}
T^\mu_\nu = diag\{\rho,-p,-p,-p\} 
\end{equation}
to give 
\begin{equation}
\rho = {1 \over 2} \dot\sigma^2 + {1 \over 2 a^2} 
\vert\nabla\sigma\vert^2 + V(\sigma) 
\end{equation}
\begin{equation}
p = {1 \over 2} \dot\sigma^2 - {1 \over 6 a^2} \vert\nabla\sigma\vert^2 - 
V(\sigma). 
\end{equation}

Plugging these expressions into Eq. (\ref{adotdot}) and using Eq. 
(\ref{hubble}) to eliminate the $\dot\sigma$ term gives 
\begin{equation}
\ddot a = -2 {\dot a^2 \over a} + 8 \pi a \left({1 \over 3 a^2} 
\vert\nabla\sigma\vert^2 + V(\sigma)\right). 
\end{equation}
Switching now to the fields $\phi$ and $\chi_i$ and to the potential in 
Eq. (\ref{potential}), 
\begin{equation}\label{adotdotbare}
\ddot a =  -2 {\dot a^2 \over a} + 8 \pi a \left({1 \over 3 a^2} 
\left(\vert\nabla\phi\vert^2 + \vert\nabla\chi_i\vert^2\right) + {\lambda 
\over 4} \left(\phi^2 - v^2\right)^2 + {1 \over 2} g^2 \phi^2 
\chi_i^2\right). 
\end{equation}
Finally we rescale all variables according to Eq. (\ref{rescalings}) and 
take a spatial average (denoted by ``$<>$") over the grid 
\begin{equation}\label{aevolution}
a'' =  -{a'^2 \over a} + {8 \pi \phi_0^2 \over a}\left<{1 \over 3} 
\left(\vert\nabla_{pr}\phi_{pr}\vert^2 + 
\vert\nabla_{pr}\chi_{i,pr}\vert^2\right) + {1 \over 4} \left(\phi_{pr}^2 
- a^2 v_{pr}^2\right)^2 + {1 \over 2} ge\ \phi_{pr}^2 
\chi_{i,pr}^2\right>. 
\end{equation}

\subsection{Initial Conditions}

Although the field equations are solved in configuration space with each 
lattice point representing a position in space, the initial conditions 
are set in momentum space and then Fourier transformed to give the 
initial values of the fields and their derivatives at each grid point. It 
is assumed that no significant particle production has occurred before 
the beginning of the program, so Minkowski space quantum fluctuations are 
used for setting the initial values of the modes.\footnote{This 
approximation is usually very good for fluctuations with momenta greater 
than $H$. For smaller momenta one can find a better approximation after 
performing the field quantization in curved space, see e.g. \cite{FKL}, 
but in our case the corresponding corrections only slightly modify the 
final result.} 

The calculation of these modes is given in detail in \cite{lattice}. The 
result is that each mode $\sigma_k$ has a probability distribution given 
by 
\begin{equation}
P(\sigma_k,\sigma^*_k) \propto e^{-2 \omega_k \sigma_k \sigma^*_k} 
\end{equation}
where $\omega_k \equiv \sqrt{k^2 + m^2}$. Separating $\sigma_k$ into a 
magnitude and a phase, the phase has a uniform probability distribution 
and the magnitude has a Rayleigh distribution 
\begin{equation}
P(\vert\sigma_k\vert) \propto \vert\sigma_k\vert e^{-2 \omega_k 
\vert\sigma_k\vert^2}. 
\end{equation}
The momentum $k$ is simply the Fourier transform variable and the mass is 
given by 
\begin{equation}
m^2_\sigma = {\partial^2 V \over \partial \sigma^2} \approx 
\left\{\begin{array}{ll} 3 \lambda \phi_0^2 &, \phi 
\\g^2 \phi_0^2&, \chi_i\end{array}\right.
\end{equation}

Taking into account the finite size of the box, $L$, and the 
discretization of the spatial points with spacing $dx$ and converting to 
the rescaled variables defined above yields for the initial magnitudes 
\begin{equation}
\vert\phi_{k,pr}\vert = {\sqrt{\lambda} L_{pr}^{3/2} \over \sqrt{2} 
dx_{pr}^3 \left(k_{pr}^2 + 3\right)^{1/4}} 
\end{equation}
\begin{equation}
\vert\chi_{ik,pr}\vert = {\sqrt{\lambda} L_{pr}^{3/2} \over \sqrt{2} 
dx_{pr}^3 \left(k_{pr}^2 + ge\right)^{1/4}} 
\end{equation}
times a Rayleigh distributed random number with standard deviation $1$. 
Note that the program values of $L$, $dx$, and $k$ are defined by the 
rescaling of $x$, and recall that $ge \equiv g^2/\lambda$. Finally the 
zero mode, which appears as a value uniformly added to all grid points at 
the beginning of the calculation, is set to $0$ for $\chi_i$ and $1$ for 
$\phi$ (since $\phi_{pr} = \phi/\phi_0$). 

In order to set the field derivatives it is necessary to know the time 
dependence of the vacuum fluctuations being considered.  In Minkowski 
space this time dependence is given simply by the term $e^{-i \omega t}$ 
which suggests 
\begin{equation}
\dot\sigma_k = -i \omega_k \sigma_k. 
\end{equation}
Converting to rescaled values of time, mass, and momentum gives 
\begin{equation}
\phi_{k,pr}' = i \phi_{k,pr} \sqrt{k_{pr}^2 + 3} 
\end{equation}
\begin{equation}
\chi_{ik,pr}' = i \chi_{ik,pr} \sqrt{k_{pr}^2 + ge} 
\end{equation}
The initial derivatives of the zero modes of $\phi$ and $\chi_i$ are set 
to $0$. 

Note that having been eliminated from the equations of motion, the only 
place $\lambda$ shows up in the calculations at all aside from the ratio 
$g^2/\lambda$ is in the $\sqrt{\lambda}$ term in the magnitude of the 
initial fluctuations. The coupling constant $g$ appears nowhere except in 
$g^2/\lambda$. Meanwhile the initial value of the field $\phi$, i.e. 
$\phi_0$, is used in the rescaling of the field and spacetime variables 
but it appears neither in the equations of motion for the fields nor in 
the amplitude of their initial fluctuations. In fact it appears in two 
places in the calculation. The first is in the evolution equation for the 
scale factor, Eq. (\ref{aevolution}). The second way the initial value of 
$\phi$ enters the calculations is subtler. We said that the zero mode of 
$\phi_{pr}'$ is initially set to $0$. At first this may seem like a poor 
choice since the field $\phi$ initially must be rolling towards $0$. In 
fact the beginning of the program is supposed to represent the end of 
inflation when the slow roll approximation is no longer valid, so why 
should $\phi_{pr}'$ be set to $0$? The answer comes from the use of the 
conformal field $\phi_{pr} \propto a \phi$. As time goes on $\phi$ is 
decreasing but $a$ is increasing, and there is a moment when these two 
balance and $\phi_{pr}'$ is momentarily $0$. By setting $\phi_{pr}'=0$ 
the program automatically sets the beginning of the calculation at this 
moment. For the potential described here this moment occurs when $\phi 
\approx .35 M_p$, so the initial conditions implicitly use the value of 
$\phi_0$. In the one place in the program where it does appear (in the 
scale factor evolution) this parameter is set to $.35$. This value is 
also useful in converting the output of the program to physically 
meaningful units. 

\subsection{The Calculational Method: Staggered Leapfrog}

The differential equations derived above are solved using a staggered 
leapfrog algorithm in which the field values and their time derivatives 
are calculated at alternating times separated by a half time-step. The 
time step is kept constant throughout the calculation. (Since the program 
uses conformal time the {\it physical} time elapsed during each time step 
changes as the program progresses.) This method is stable for second 
order equations involving no first time derivatives such as the field 
equations we use, see e.g. \cite{leapfrog}. However, extra care must be 
taken in solving the equation for the scale factor, Eq. 
(\ref{aevolution}), since this does contain a first derivative term $a'$. 
A naive calculation using the leapfrog algorithm as described above would 
mean that $a''$ would be calculated at time $\tau$ as a function of 
$a(\tau)$ and $a'(\tau - d\tau/2)$. There is a solution for this, 
although in practice the evolution of $a$ is so slow and smooth that this 
problem makes no practical difference. We avoid this problem, though, by 
using the two following equations\footnote{G.F. would like to thank 
Julian Borrill, who suggested this solution.} 
\begin{equation}\label{adotadvance}
a'(\tau + d\tau/2) \approx a'(\tau - d\tau/2) + d\tau\ a''(\tau) 
\end{equation}
\begin{equation}\label{adotaverage}
a'(\tau) \approx {1 \over 2} \left(a'(\tau + d\tau/2) + a'(\tau - 
d\tau/2)\right). 
\end{equation}
Solving these simultaneously, using Eq. (\ref{aevolution}) for $a''$, 
gives 
\begin{equation}
a'(t + d\tau/2) \approx -a'(\tau - d\tau/2) + {2 a \over d\tau} \left(-1 
+ \sqrt{1 + 2 {a'(\tau - d\tau/2) \over a} d\tau + {f \over a} 
d\tau^2}\right) 
\end{equation}
where 
\begin{equation}
f(a(\tau)) = {8 \pi \phi_0^2 \over a}\left<{1 \over 3} 
\left(\vert\nabla\phi_{pr}\vert^2 + \vert\nabla\chi_{i,pr}\vert^2\right) 
+ {1 \over 4} \left(\phi_{pr}^2 - a^2 v_{pr}^2\right)^2 + {1 \over 2} ge\ 
\phi_{pr}^2 \chi_{i,pr}^2\right>. 
\end{equation}
Since Eq. (\ref{aevolution}) needs to be solved only once per time step 
this correction involves virtually no added computational time.


\newpage

\begin{thebibliography}{99}

\bibitem{Kirzhnits} D.A.~Kirzhnits, JETP Lett. {\bf 15}, 529 (1972);
D.A.~Kirzhnits and A.D.~Linde, Phys. Lett. {\bf 42B}, 471 (1972). 

\bibitem{Weinb}
S.~Weinberg, Phys. Rev. {\bf D9}, 3320 (1974); L.~Dolan and R.~Jackiw, 
Phys. Rev. {\bf D9}, 3357 (1974); D.A.~Kirzhnits and A.D.~Linde, Sov. 
Phys. JETP {\bf 40}, 628 (1974). 

\bibitem{Kirzhnits2} D.A.~Kirzhnits and A.D.~Linde, Ann. Phys. {\bf  
101}, 195 (1976). 

\bibitem{book}
A. D. Linde, {\em Particle Physics and Inflationary Cosmology} (Harwood, 
Chur, Switzerland, 1990) 

 
\bibitem{infl} A.H. Guth, Phys. Rev. {\bf D23}, 347 (1981); A.D.
Linde, Phys. Lett. {\bf 108B}, 389 (1982); A. Albrecht and P.J. 
Steinhardt, Phys. Rev. Lett. {\bf 48}, 1220 (1982). 

\bibitem{Shap} V.A. Kuzmin, V.A. Rubakov and M.E.Shaposhnikov, Phys.
Lett. {155B} (1985) 36; M.E. Shaposhnikov, JETP Lett. {\bf 44} (1986) 
465; Nucl. Phys. {\bf B287} (1987) 757; Nucl. Phys. {\bf B299} (1988) 
797. 


\bibitem{KLS} L. Kofman, A. Linde, and A. A. Starobinsky,
Phys. Rev. Lett. {\bf 73}, 3195 (1994); L. Kofman, A. Linde, and A. A. 
Starobinsky, Phys. Rev. D {\bf 56}, 3258 (1997). 
 

\bibitem{ptr1}
L. Kofman, A. Linde, and A. A. Starobinsky, Phys. Rev. Lett. {\bf 76}, 
1011 (1996). 

\bibitem{ptr2}
I.~I.~Tkachev, Phys. Lett. {\bf B376}, 35 (1996). 

\bibitem{bubble} S. Khlebnikov, L. Kofman, A. Linde, and I. Tkachev,
 Phys.
Rev. Lett. {\bf 81}, 2012 (1998), hep-ph/9804425. 

\bibitem{Copeland} A.~Rajantie and E.~J.~Copeland,
 ``Phase transitions from preheating in gauge theories,'' hep-ph/0003025. 
 

\bibitem{LS} D.H. Lyth and  E.D. Stewart, Phys. Rev. Lett. {\bf 75},
201 (1995).  

\bibitem{strings} I. Tkachev, S. Khlebnikov, L. Kofman, and A. Linde,  Phys.
Lett. {\bf B440}, 262-268 (1998), hep-ph/9805209. 

\bibitem{kawasaki} S. Kasuya, M. Kawasaki, 
Phys. Rev. {\bf D61}, 083510 (2000),  hep-ph/9903324. 


\bibitem{lattice} S. Khlebnikov, and I. Tkachev,
  Phys. Rev. Lett. {\bf 77}, 219 (1996), 
hep-ph/9603378; S. Khlebnikov and I.~I.~Tkachev, Phys. Rev. Lett. {\bf 
79}, 1607 (1997). 

 

\bibitem{kolb}
E. W. Kolb and M. S. Turner, {\em The Early Universe} (Addison-Wesley, 
Redwood City, California, 1990). 


 

\bibitem{AL92} A.D. Linde,   Nucl. Phys. {\bf B372}, 421 (1992). 

\bibitem{coleman} S. Coleman, Phys. Rev. {\bf D15} (1977) 2929.

\bibitem{AL83}  A.D. Linde,  Nucl.Phys. {\bf B216},  421 (1983).

\bibitem{zeld} Ya.B.~Zeldovich, I.Yu.~Kobzarev, and L.B.~Okun,
Phys. Lett. {\bf 50B}, 340 (1974). 

\bibitem{LR} A.D. Linde and A. Riotto, 
Phys. Rev. D {\bf 56}, 1841  (1997),  hep-ph/9703209; C. Contaldi, M. 
Hindmarsh, J. Magueijo, Phys. Rev. Lett. {\bf 82}, 2034 (1999), 
astro-ph/9809053; R.A. Battye, J. Weller, Phys. Rev. {\bf D61}, 043501 
(2000), astro-ph/9810203; R. A. Battye, J. Magueijo, J. Weller, 
astro-ph/9906093. 

\bibitem{KKLP} R.~Kallosh, L.~Kofman, A.~Linde and A.~Van Proeyen,
``Gravitino production after inflation,'' hep-th/9907124. 

\bibitem{GTR}  G.~F.~Giudice, I.~Tkachev, and A.~Riotto,
JHEP {\bf 9908}, 009 (1999), hep-ph/9907510. 

\bibitem{FKL}  G.~Felder, L.~Kofman, and A.~Linde, 
``Gravitational particle production and the moduli problem,''  JHEP, {\bf 
02}, 027 (2000), hep-ph/9909508. 


\bibitem{leapfrog} K.C.B. New, K. Watt, and C.W. Centrella,
Stable Three Level Leapfrog Integration in Numerical Relativity, Phys. 
Rev. {\bf D58}, 064022 (1998), gr-qc/9801110. 


\end{thebibliography}
\end{document}